\documentclass[11pt,twoside,a4paper]{article}
\usepackage{siunitx}
\usepackage{amsmath}
\usepackage{hyperref}
\usepackage{verbatim}
\usepackage{amsfonts}
\usepackage{graphicx}
\usepackage{subcaption}
\usepackage{floatpag}
\usepackage{amssymb}
\usepackage{epstopdf}
\usepackage{epsfig}
\usepackage{pdfpages}
\usepackage{soul}
\usepackage{siunitx}
\usepackage{cite}
\usepackage[left=1in, right=1in, top=1in, bottom=1in]{geometry}

\begin{document}

\title{Optical analogues of black-hole horizons}

\author{Yuval Rosenberg\\
{\small Department of Physics of Complex Systems,}\\
{\small	Weizmann Institute of Science, Rehovot 761001, Israel.}}

	\maketitle

\begin{abstract}
Hawking radiation is unlikely to be measured from a real black-hole, but can be tested in laboratory analogues. It was predicted as a consequence of quantum mechanics and general relativity, but turned out to be more universal. A refractive index perturbation produce an optical analogue of the black-hole horizon and Hawking radiation that is made of light. We discuss the central and recent experiments of the optical analogue, using hands-on physics. We stress the roles of classical fields, negative frequencies, 'regular optics' and dispersion. Opportunities and challenges ahead are shortly discussed. 
\end{abstract}

\section{Introduction}
Analogue systems are used to test theories of predicted phenomena that are hard to observe directly \cite{georgescu2014quantum_s}. Hawking \cite{hawking1974black, hawking1975particle} predicted black-holes to thermally radiate from the quantum vacuum. Its measurement is unlikely, having a temperature inversely related to the black-hole mass. Stellar and heavier black-holes' radiation is much weaker than the cosmic microwave background (CMB) fluctuations \cite{CMB_pedestrians,HawkingTemp}. Hypothetical microscopic primordial black-holes might emit significant Hawking radiation \cite{halzen1991gamma}, but it is highly model-dependent, with rates seriously bounded by the lack of its observation \cite{alexandreas1993new, fichtel1994search, linton2006new}. 

Volovik and Unruh \cite{volovik2003universe, unruh1981experimental} have shown that a transonic fluid flow is analogous to the space-time geometry surrounding a black-hole, and should emit sound waves analogous to Hawking radiation. 

Wave kinematics control the Hawking process, which appears in the presence of an effective horizon \cite{visser1998acoustic, visser2003essential}. Black-hole dynamics and Einstein equations of gravity are not an essential requirement. The analogue of the event horizon is the surface separating the subsonic and supersonic flows. The surface gravity determines the strength of Hawking radiation and is analogous to the flow acceleration at the horizon. The horizon should also last long enough such that the modes of Hawking radiation will have well defined frequencies \cite{visser2003essential}. 

Many analogues have been proposed \cite{barcelo2011analogue, barcelo2019analogue} with realisations using water waves \cite{rousseaux2008observation, jannes2011experimental, weinfurtner2011measurement, euve2015wave, euve2016observation, torres2017rotational}, Bose-Einstein condensates (BEC) \cite{lahav2010realization, steinhauer2014observation, steinhauer2016observation, de2019observation}, and optics \cite{philbin2008fiber, belgiorno2010hawking, rubino2012negative,elazar2012all,nguyen2015acoustic, bekenstein2015optical, bekenstein2017control, drori2019observation}. The analogues gave new perspectives in both gravity and the analogue systems \cite{barcelo2011analogue}, extending Hawking's theory to wave dynamics in moving media. The universality of the Hawking effect shows it persists in real-world scenarios, where complicated dynamics replace simplified and possibly fine-tuned theories. Hawking's original derivation seems questionable, since the radiation originates from infinitely high frequencies, neglecting unknown high energy physics (known as the 'trans-Planckian problem'). The microscopic physics of the analogue systems is well known, allowing to test the role of high frequencies in Hawking predictions \cite{jacobson1991black,unruh1995sonic}.

This paper discusses the optical analogues that use a refractive index perturbation to establish an artificial black-hole horizon. Extensive reviews of analogue gravity exist \cite{barcelo2011analogue, barcelo2019analogue}, but they pay little attention to optical analogues and substantial progress was made after their publication. Additional useful resources include Refs. \cite{daniele2018hawking,faccio2013analogue, leonhardt2010essential,unruh2007quantum}. This paper aims to present the main experiments that transformed the optical analogues from simple ideas to established reality. It aims to do so in the light of the recent developments, but in terms of hands-on physics. We believe that experiments reach the state where their data may lead the way for new questions and discoveries in the physics of Hawking radiation.

Section 2. briefly describes early proposals for optical analogues using slow light and why they could not materialise. Transformation optics is also mentioned, where stationary dielectrics are viewed to change the spatial parts of the metric.
Section 3. explains the standard theory of optical analogues discussed in this paper and the first demonstration of optical horizons. Further measurements of the frequency shift at group velocity horizons are briefly discussed, and additional related work is mentioned. 
Section 4. focuses on attempts in bulk optics rather than optical fibres, stressing the roles of group-velocity and phase-velocity \cite{milonni2004fast} horizons. 
Section 5. discusses the first measurements of negative frequencies in optics, a phenomenon closely related to the Hawking effect. Its theory is also presented and shortly explained.
Section 6. considers additional interpretations of the optical horizon. Theory and experiments are shown to directly relate the effect to cascaded four-wave mixing, but analysis in the time domain seems to be unavoidable. The use of temporal analogues of reflection and refraction, and numerical solutions are also mentioned. 
Section 7. discusses the first demonstration of stimulated Hawking radiation in an optical analogue, and lessons learnt from it. 
Section 8. concludes the paper with a brief outlook to the future.

\section{Early attempts}

Fresnel's drag \cite{Fresnel} was an ether-based theory of light propagation in transparent media, 'confirmed' by Fizeau \cite{fizeau}. The drag effect is just a relativistic velocity addition\cite{rindler2006relativity}, but Fresnel's wrong theory was based on correct intuition of velocity addition. The ether was replaced by the space-time geometry and the quantum vacuum, which continued to have an intimate connection to moving media and produce puzzling phenomena \cite{hawking1974black,hawking1975particle, fulling1973nonuniqueness,davies1975scalar,unruh1976notes}. Analogue gravity took this connection a step further, but despite theoretical progress in the 1990's, no practical realization of analogue black-holes was suggested at that time \cite{barcelo2011analogue}.

Technology drove ideas in the right directions (optics and Bose-Einstein condensates) around the year 2000 \cite{leonhardt1999optics,leonhardt2000slow_PRL, leonhardt2002slow_nature, garay2000sonic,garay2001sonic}. In optics, Leonhardt and Piwnicki \cite{leonhardt1999optics,leonhardt2000slow_PRL, leonhardt2002slow_nature} suggested to slow down light such that its medium could be moved in super-luminal velocities and form a horizon. These ideas used the technology of 'slow light' \cite{hau1999Slow_light_1,kash1999ultraslow}, where incredibly low group velocities are produced using electromagnetically induced transparency (EIT, see Fig.~\ref{fig3}a). However, the phase velocity of 'slow light' is fast, preventing the crucial formation of a phase velocity horizon (where the media moves at the phase velocity of light. See below) \cite{unruh2003slow}. Another problem with realising these ideas is the narrow bandwidth of light that can be slowed \cite{unruh2003slow}, and severe absorption around it \cite{milonni2004fast}. The inevitable conclusion was to move the medium in relativistic velocities.

Despite missing key concepts of Hawking radiation, these ideas have pushed analogue gravity forward and beyond the scope of relativity (or relativitists). They showed that an analogue black-hole metric can be made in optics. Similarly, stationary dielectrics mimic spatial geometries. This interpretation inspired transformation optics, and the development of meta-materials technology extended the range of practical geometries \cite{leonhardt2006optical, pendry2006controlling, leonhardt2006general, chen2010transformation, leonhardt2010geometry, xu2015conformal}. 

\section{Changing a reference frame}

\begin{figure}[t!]
	\centering\includegraphics[width=0.9\textwidth]{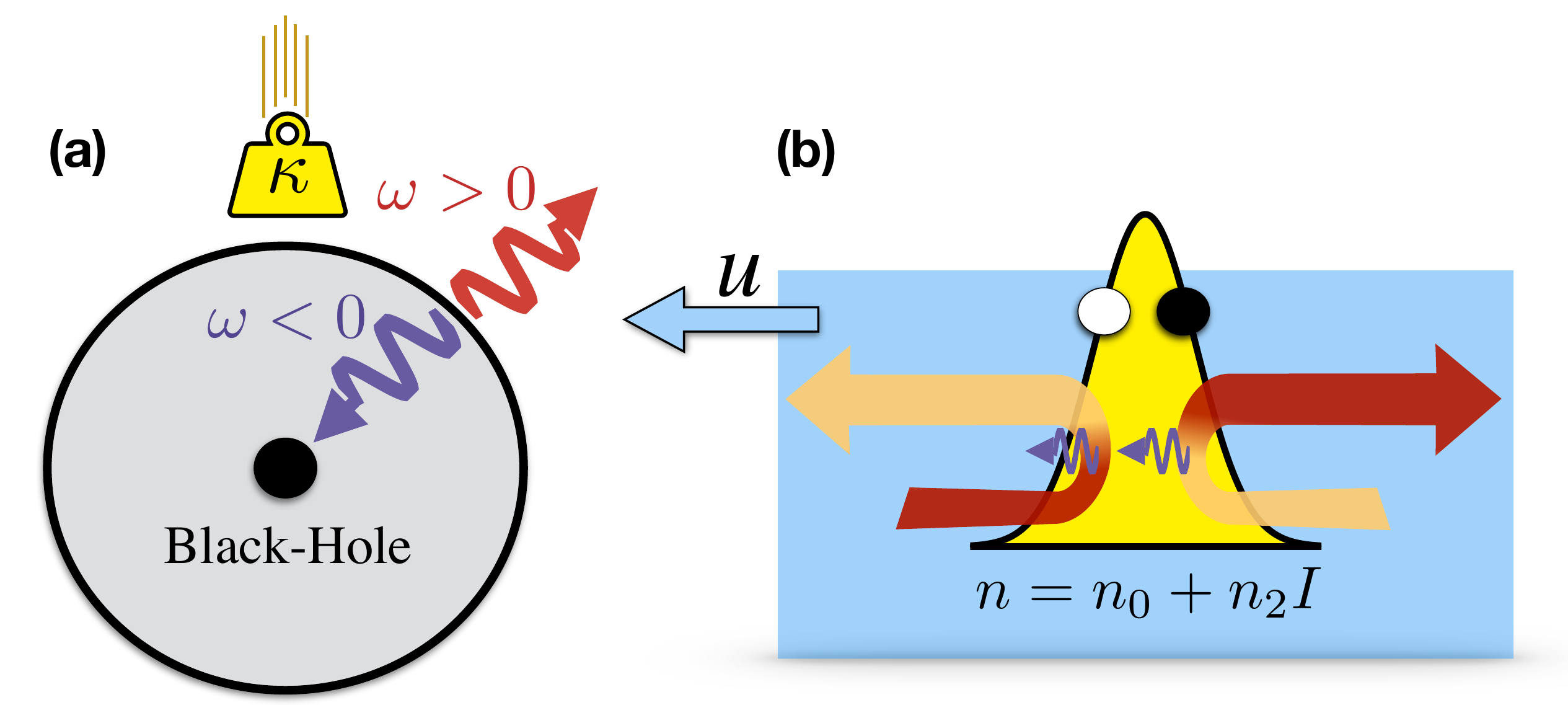}
	
	\caption{Main optical analogue concepts. \textbf{(a)} A black-hole emit Hawking radiation from its event horizon: positive norm waves escape outside, while negative norm waves drift towards its singularity. The surface gravity $\kappa$ represented by a falling weight determines the radiation's strength. \textbf{(b)} The medium seems to flow in the pulse co-moving frame with its group velocity $u$. A pair of black-hole and white-hole horizons are formed due to nonlinear refraction index being proportional to the pulse local intensity. Positive frequency probe light is shifted, and negative frequency waves are formed at both horizons according to the dispersion relation (see Fig.~\ref{fig2}). The magnitude of the Hawking effect is determined by the pulse steepness, being analogous to the surface gravity at the horizon (see text).}
	\label{fig1}
\end{figure}

The groups of Leonhardt and K{\"o}nig \cite{philbin2008fiber} started a new approach that follows a simple idea \cite{Leonhardt2005Invent}: light itself travels at the speed of light. A light pulse (also denoted pump) travels in a dielectric medium with group velocity $u$ . In the reference frame co-moving with the pulse ('the co-moving frame'), the medium seems to flow in the opposite direction with velocity magnitude $\left|u\right|$ (Fig.~\ref{fig1}b). Probe light of a different frequency differ in velocity due to dispersion. The nonlinear Kerr effect \cite{boyd2003nonlinear} slows down the probe upon interaction with the pulse\footnote{This has a similar effect to the flow acceleration in the fluid analogues, changing the probe velocity relative to the flow.} -- the refractive index change as $n=n_0+n_2 I$, where $n_0$ is the linear refractive index, $n_2$ is a parameter being typically  $10^{-16}\text{cm}^2/\text{W}$ \cite{boyd2003nonlinear}, and $I$ is the light intensity. A group velocity horizon forms where the probe group velocity obey $v_g=\left|u\right|$, blocking the probe from further entering the pump. A black-hole horizon is formed in the leading end of the pump, where the flow is directed into the pulse. Its time reversal, a white-hole horizon with outward flow, is formed in the pump trailing end \cite{philbin2008fiber, belgiorno2011dielectric}.

If the pump is slowly varying \cite{visser2003essential}, the probe co-moving (and Doppler shifted) frequency, 
\begin{equation}\label{dopplerEq}
\omega'=\gamma \left(1-n \frac{u}{c}\right) \omega,
\end{equation}
is conserved. Here $\gamma=\left(1-u^2/c^2\right)^{-1/2}$ is the Lorentz factor, $c$ is the vacuum speed of light, $n$ is the refraction index, and $\omega$ is the probe frequency in the laboratory frame. Photon pairs of Hawking radiation are produced at each horizon: one with positive $\omega'$, and its negative partner with $-\omega'$ \cite{leonhardt2010essential, leonhardt2012analytical}. For positive $\omega$, $\omega'$ is positive only if the probe phase velocity, $v_\phi$, is greater than $|u|$. A phase velocity horizon forms where $v_\phi=\left|u\right|$ so $\omega'=0$. The Hawking radiation outgoing from the horizon mix incoming radiation of positive and negative frequencies \cite{leonhardt2010essential}. Quantum mechanically, it is described by a Bogoliubov transformation 
\begin{equation}\label{BogoliubovEq}
\hat{b}_\pm=\alpha \hat{a}_\pm + \beta \hat{a}^\dagger_\mp \quad;\quad |\alpha|^2-|\beta|^2=1,
\end{equation}
where the sign of the operators equals the sign of $\omega'$. Time dependent annihilation and creation operators for incoming modes, $\hat{a}_\pm,\hat{a}^\dagger_\mp$ mix to form annihilation operators for outgoing modes, $\hat{b}_\pm$. This extracts metric energy (pump energy) to amplify the radiation, thus spontaneously creating outgoing radiation from incoming vacuum \cite{leonhardt2010essential, drori2019observation}. The transformation parameters, $\alpha$ and $\beta$, give the flux of Hawking radiation. When neglecting dispersion, the radiation effective temperature is proportional to the analogue of the surface gravity -- the steepness of the pulse (giving the probe velocity gradient in the co-moving frame \cite{visser2003essential, leonhardt2010essential}, which is related to the refractive index (and pulse intensity) gradient or rate of change \cite{philbin2008fiber,leonhardt2010essential,belgiorno2011dielectric}).

\begin{figure}[t!]
	\centering\includegraphics[width=0.9\textwidth]{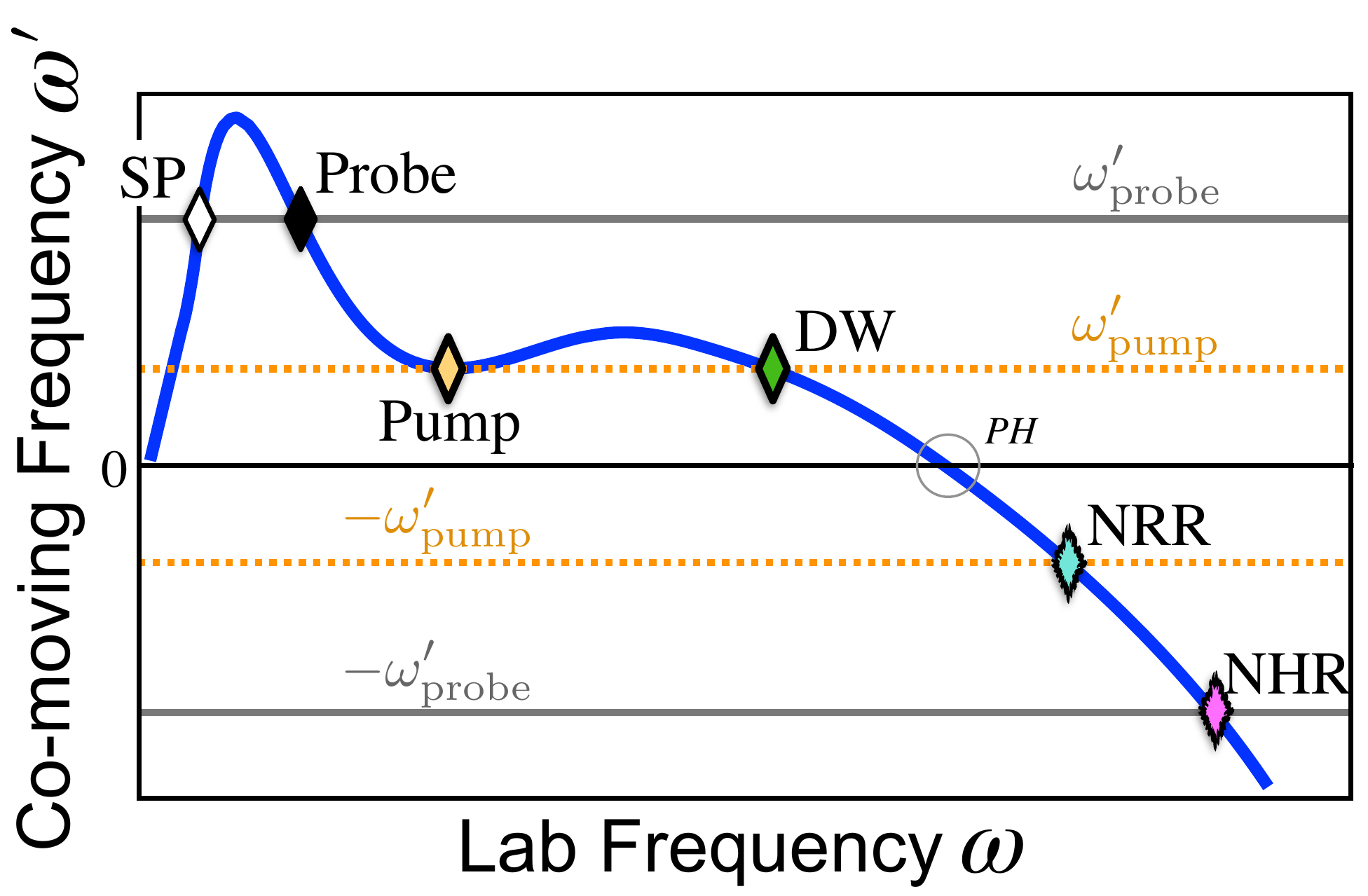}
	
	\caption{Schematic Doppler curve of the fibre optic analogue with phase matching conditions. Frequency $\omega'$ at the reference frame co-moving with the pump pulse is plotted against the laboratory frequency $\omega$ (Eq.~\ref{dopplerEq}). The pump pulse is in a local minima, where the dispersion is anomalous. The phase matching condition of dispersive waves (DW) corresponds to conservation of $\omega'_\text{pump}$ (upper dotted line). Negative frequency resonant radiation (NRR) conserve $-\omega'_\text{pump}$ (lower dotted line. See also Fig.~\ref{fig3}d,\ref{fig3}f). Probe light (black-diamond) is red-shifted (white diamond) at black-hole horizon, conserving its co-moving frequency $\omega'_\text{probe}$ (upper solid line). At the white-hole horizon the probe (now white diamond) is blue-shifted (black diamond), as seen also in Fig.~\ref{fig3}b. At the local maxima the group-velocity matches that of the pump, $u$. Negative Hawking radiation (NHR) conserve $-\omega'_\text{probe}$ (lower solid line. See also Fig.~\ref{fig3}f). The phase horizon (PH), where the phase velocity equals $u$, corresponds to $\omega'=0$. 
	}
	\label{fig2}
\end{figure}

In \cite{philbin2008fiber}, an optical fibre called a photonic crystal fibre (PCF) provided both the desired dispersion (Fig.~\ref{fig2}) and nonlinearity. Its unique structure guided light in a very small region -- the fibre's core \cite{russell2003photonic, Agrawal_NLFO}. This increased the light intensity and the fibre's nonlinear response. Its nonlinear parameter was $\gamma\left(\omega_0\right) = \omega_0 n_2 /c A_\text{eff} = 0.1 \text{W}^{-1}\text{m}^{-1}$ at $\omega_0$ corresponding to $\SI{780}{\nano\meter}$ wavelength, where $A_\text{eff}$ is the fibre effective mode area and $\epsilon_0$ is the vacuum permittivity \cite{Agrawal_NLFO}. The fibre's structure was engineered to change its dispersion relation, $n(w)$, to have two points with matching group velocities \cite{Agrawal_NLFO}: one in a normal dispersion region, where $n(w)$ is increasing; and another in an anomalous dispersion region, where $n(w)$ is decreasing. The anomalous dispersion included the pump spectra, generated by a Ti:Sapphire mode-locked laser. Self-phase modulation (SPM) due to the nonlinear refraction index counteracted the anomalous dispersion and formed stable solitons \cite{Agrawal_NLFO}.

The solitons in \cite{philbin2008fiber} created horizons. Continuous wave (CW) probe was added to the fibre, being red-detuned from the frequency of matching group velocity (white diamond in Fig.~\ref{fig2}). The fast probe mainly interacted with the pump trailing end, gradually slowed down, and blue-shifted (frequency up-conversion, see Fig.~\ref{fig3}b). This pump-probe interaction is known as cross phase modulation (XPM), and induces chirp that depends on the pump pulse shape \cite{Agrawal_NLFO}. If $\delta n=n_2 I$ is large enough to form a horizon, the shifting probe become slower than the pulse and they separate. In the co-moving frame the probe is reflected (Fig.~\ref{fig1}b) while conserving its co-moving frequency $\omega'_\text{probe}$, but not $\omega$ (Fig.~\ref{fig2}). This reflection demonstrated the existence of optical horizons \cite{philbin2008fiber}. 

Philbin et.al. \cite{philbin2008fiber} used $\SI{70}{\femto\second}$ duration FWHM (full width at half maximum), $\SI{800}{\nano\meter}$ carrier wavelength pulses with peak power of about $\SI{50}{\watt}$. The PCF was $\SI{1.5}{\meter}$ long with core diameter smaller than $\SI{2}{\micro\meter}$. The CW probe was tunable around $\SI{1500}{\nano\meter}$ wavelength with $\SI{100}{} - \SI{600}{\micro\watt}$ power. Pump power was kept low to maintain a stable single soliton and reduce high-order nonlinearities, such as the Raman effect \cite{Agrawal_NLFO}. Self-steepening was hoped to realize high Hawking temperatures. 

This set-up achieved a mere $10^{-3}$ percent efficiency for probe shifting. Only $10^{-4}$ of the CW probe power was expected to interact with the pump over the $\SI{1.5}{\meter}$ fibre, and another order of magnitude reduction was attributed to tunnelling of the probe through the narrow pump barrier \cite{philbin2008fiber}. While clearly showing probe blue-shifting at the white-hole horizon, this set-up could not produce detectable negative frequency partners. It was unclear whether they should form around the phase velocity horizon, where $\omega'=0$, or around the supported frequency $-\omega'_\text{in}$ (Fig.~\ref{fig2}).

Choudhary and k{\"o}nig \cite{choudhary2012efficient} reported probe red-shifting and studied its tunnelling. They used similar experimental parameters to \cite{philbin2008fiber}, with tuneable $\SI{50}{\femto\second}$ pump pulses and visible CW probe at $\SI{532}{\nano\meter}$ (to avoid the difficulty of synchronisation). Tunnelling was minimal for probe detuning up to twice the soliton bandwidth (relative to the matching frequency), but limited pump-probe interaction kept the conversion efficiency small. Tartara \cite{tartara2012frequency} derived both pump and probe from the same $\SI{105}{\femto\second}$ source and propagated them in a $\SI{1.1}{\meter}$ fibre, demonstrating conversion efficiencies of tens of percents. An optical parametric amplifier produced the tuneable probe.

The Raman effect cause the pump to decelerate, and was shown to only slightly change the shifted probe spectra \cite{robertson2010frequency}. Shifting of dispersive waves \cite{wang2015optical,bendahmane2015observation} and trapping the probe light \cite{nishizawa2002pulse, nishizawa2002characteristics, gorbach2007light ,hill2009evolution, wang2013soliton} were also related to the Raman effect. An optical 'black-hole laser' that use both the white and black-hole horizons was predicted and analysed \cite{corley1999black , leonhardt2007black , faccio2012optical, gaona2017theory, bermudez2018resonant}. The horizon dynamics was also related to the formation of optical rogue waves and champion solitons \cite{solli2007optical, demircan2012rogue , demircan2014rogue, pickartz2016adiabatic},  and the ability to form all-optical transistors \cite{demircan2011controlling}. Similar 'front-induced transitions' were studied in other areas of photonics \cite{gaafar2019front}, and the quest for measuring analogue Hawking radiation continued.

\section{Horizonless emissions}
Faccio's group realized optical phase velocity horizons, without a group velocity horizon, in bulk optics \cite{belgiorno2010hawking, rubino2011experimental}. They constructed pulses of super-luminal group velocities, by using their three dimensional nature \cite{faccio2007conical}. A pulsed Bessel beam is one such example. It is composed of infinitely many plane waves on a cone with a constant angle $\theta$ with the propagation axis. It can be made with an axicon lens \cite{faccio2007conical}. At the apex of the cone, the plane waves interfere to form a bright spot that moves with super-luminal velocity that depends on $\theta$. Belgiorno et al. \cite{belgiorno2010hawking} used super-luminal pulses that travelled as narrow and powerful filaments inside fused silica. Looking perpendicularly to their propagation direction, they detected radiation around the phase velocity horizon -- where the radiation phase velocity matched the pulses group velocity (Fig.~\ref{fig3}c). They could not measure along the propagation direction, because vast radiation was produced by the powerful pulses there (peak intensity was as high as $10^{13}\text{W}/\text{cm}^2$ centred at $\SI{1055}{\nano\meter}$ wavelength). 

Spurious radiation was a key issue even for observation at 90 degrees. Special care was taken to tune the radiation into spectral windows of minimal noise. Correctly identifying all the signals in such a set-up is another major challenge. Concerns were raised \cite{schutzhold2010comment, unruh2012hawking, liberati2012quantum} and in-depth analysis related the radiation to a horizonless super-luminal perturbation, possibly linked to Hawking radiation \cite{petev2013blackbody, finazzi2014spontaneous}. This work stressed the importance of a blocking group velocity horizon to the Hawking effect. Thermal Hawking radiation was predicted for super-luminal pulses in materials with linear dispersion at low energies (like diamond) \cite{petev2013blackbody, finazzi2014spontaneous}, but is yet to be measured. 

Leonhardt and Rosenberg \cite{leonhardt2019cherenkov} related the emission in \cite{belgiorno2010hawking} to the physics of Cherenkov radiation\footnote{Not to be confused with dispersive waves \cite{akhmediev1995cherenkov}} in a surprising way. The pulse was modelled as a super-luminal light bullet \cite{silberberg1990collapse}, and found to behave like a moving magnetic dipole that is predicted to emit Cherenkov radiation with a discontinuous spectrum \cite{frank1942doppler, frank1984vavilov}. This discontinuity is exactly at the phase horizon -- where the effective dipole velocity equals the phase velocity of light and $\omega'=0$ (Fig.~\ref{fig2}). Interferences turn the discontinuity into a peak if the dipole is an extended object, like the light bullet.

\section{The role of negative frequencies}
Negative frequencies are an integral part of Hawking radiation and can also appear as dispersive waves. Solitons emit dispersive waves (also known as resonant radiation) at a shifted frequency when disturbed by higher order dispersion. Non-dispersive three dimensional light bullets emit similar radiation \cite{faccio2007conical}. Both emissions can be viewed to originate from self-scattering of the pulse by its nonlinear refractive index barrier. Conservation of momentum dictates the emitted frequency through a phase matching condition \cite{dudley2006, Agrawal_NLFO}. In the reference frame co-moving with the stable pulse, this condition becomes a conservation of energy, given by the pulse co-moving frequency, $\omega'_\text{pulse}$ (Fig.~\ref{fig2}). The frequency shifting at optical horizons extends this picture to the scattering of probe light. The prediction of negative Hawking modes suggested that negative dispersive waves should similarly appear. Rubino et al. \cite{rubino2012negative} had measured this negative-frequency resonant radiation (NRR) after it was neglected and disregarded for a long time. This required very rapid (non-adiabatic) temporal changes of the pulse envelope.

Intense pulses created NRR through steep shocks in two ways \cite{rubino2012negative}: $\SI{7}{\femto\second}$ high order solitons \cite{Agrawal_NLFO} of about $\SI{300}{\pico\joule}$ energy and $\SI{800}{\nano\meter}$ carrier wavelength were placed in $\SI{5}{\milli\meter}$ PCFs; and pulsed Bessel beams of $\SI{60}{\femto\second}$ duration, about $\SI{20}{\micro\joule}$ energy and $\SI{800}{\nano\meter}$ carrier wavelength were sent through a $\SI{2}{\centi\meter}$ long bulk calcium fluoride ($\text{CaF}_2$), as seen in Fig.~\ref{fig3}d. Despite NRR was seen both in optical fibres and in bulk, further work was needed to exclude other possible mechanisms for the effect -- contributions due to high order spatial modes, and possible interactions between the complex pulse and additional radiation.

Rubino et al. \cite{rubino2012soliton} later used a relativistic scattering potential to directly explain NRR formation, and Petev et al. \cite{petev2013blackbody} further related it to Hawking radiation. Conforti et al. \cite{conforti2013negative} extended the theory to pulses in materials of normal dispersion (where the pulse experience dispersive broadening) and dominating second order nonlinearity, which effectively creates a nonlinear refractive index. The same conditions for phase matching and non-adiabatic temporal evolution were found. Analytic derivation of the NRR \cite{conforti2013interaction} directly related it to interactions between fields and their conjugates (or the mixing of positive and negative frequency modes). It further strengthened the connection between NRR and Hawking radiation, which originate from a Bogoliubov transformation that mixes creation and annihilation operators \cite{leonhardt2010essential, leonhardt2012analytical}. McLenaghan and K{\"o}nig \cite{mclenaghan2014few} studied NRR for different PCF lengths and input chirps, and inferred UV propagation loss of $\sim 2\text{dB/mm}$. We stress that the negative frequency (and negative norm) of the Hawking modes beyond the horizon is key to the Hawking amplification process \cite{leonhardt2010essential}, extracting energy from the background metric in accordance to a Bogoliubov transformation.

\section{'It's just optics'}
Hawking radiation is a universal geometric effect that emerge due to conversion of modes at a horizon, regardless of the microscopic physics that create the background space-time geometry \cite{visser2003essential}. The same (generalized) derivation applies for both astrophysical black-holes and analogue systems \cite{visser2003essential, philbin2008fiber,finazzi2013quantum, jacquet2015quantum, bermudez2016hawking, linder2016derivation, jacquet2018_Thessis,jacquet2019analytical}. This established universality proved insightful for both gravity and optics (arguably solving the trans-Planckian problem and discovering negative frequencies in optics are two examples \cite{barcelo2011analogue, barcelo2019analogue}). However, some researchers from both the optics and gravity (or quantum field theory on curved space-times) communities are still not comfortable with analogue Hawking radiation, claiming that the effect is 'just optics' \footnote{The author have personally encountered such allegations from both communities, and believe that they partially originate from not separating 'gravity', predicted to create the black-hole event horizon, and 'kinematics', which are responsible for the Hawking effect once a horizon is formed \cite{visser2003essential}.}. Such claims do not undermine (analogue) Hawking radiation, but complete our understanding of it. It's also possible to directly explain the effect using the underlying microscopic physics -- unknown quantum gravity for real black-holes, and nonlinear optics for the optical analogues.

The classical dynamics at optical horizons can be captured using numerical solutions that take into account the entire electric field (including negative frequencies), the dispersion relation and all nonlinearities \cite{Agrawal_NLFO, amiranashvili2016hamiltonian, bermudez2016propagation, Raul2020inpress}. These are extensions and alternatives for solving the usual generalised nonlinear Schr{\"o}dinger equation (GNLSE), which is used extensively in standard descriptions of ultrafast nonlinear fibre optics \cite{Agrawal_NLFO, dudley2006, skryabin2010colloquium}. 

Another approach directly relates the phenomena to discrete photonic interactions. It uses cascaded four-wave mixing of discrete spectra, and takes the continuum limit for comparison \cite{webb2014nonlinear, erkintalo2012cascaded}. A CW probe and a pair of beating quasi-CWs (of $\SI{1}{\nano\second}$ duration) positioned symmetrically around the pump central frequency were being mixed continuously (Fig.~\ref{fig3}e). At each step of the cascaded process, the probe shifted by the pair's detuning, and all three generated an equidistant frequency comb. A resonant amplification was produced at the frequency corresponding to conservation of $\omega'_\text{probe}$. Similar amplification appeared at the dispersive wave resonance, at $\omega'_\text{pump}$. The experiments used low cascade orders ($\text{n}=5-7$ for the probe shift and about $\text{n}=15$ for the dispersive waves) and were supported by numerical analysis. This picture shows how energy is being transferred from the pump to the shifting probe, by effectively absorbing pump photons of one frequency and emitting at another. The cascaded four wave mixing also details the back reaction mechanism, where the pump undergoes spectral recoil. 

The studies \cite{webb2014nonlinear, erkintalo2012cascaded} addressed the relevant phase matching conditions, but the efficiency of the cascaded process was found only in \cite{webb2014efficiency}, for dispersive waves. Remarkably, it showed that the known concepts from horizon physics are crucial ingredients even when the pump is made of beating quasi-CWs that form a frequency comb: the mixing was analysed using (temporal) soliton fission dynamics, being efficient only when the frequency spacing between the CWs effectively generated compressing (non-adiabatic) high order solitons. The frequency-domain analysis was intractable for high cascade orders. The maximally compressed beat cycle corresponded to about $\SI{102}{\femto\second}$ duration FWHM in the efficient regime, of high cascading orders. Conversion efficiencies of $10^{-4}$ over a $\SI{100}{\meter}$ fibre were reported.

The generation of NRR in media with quadratic nonlinearity was also related to a cascading process \cite{conforti2013negative}, which must also be the case for cubic materials. However, negative frequencies were not demonstrated yet using a beating CW pump. It would be interesting to test how the cascading mixing behaves as it reaches negative frequencies, past the phase velocity horizon. 

Temporal analogues of reflection and refraction \cite{plansinis2015temporal, plansinis2018cross} are also used to explain the frequency shifts at optical horizons. It compares the frequency change due to XPM to the wave number change during refraction. The reflection at the horizon is analogous to total internal reflection. The tunnelling through the refractive index barrier is compared to frustrated total internal reflection, where an evanescent wave extends beyond the pulse and tunnels through. The optical horizon can be seen as a temporal beam splitter for probe light (acting also as an amplifier when considering negative frequencies as well). 

Comparing the mode mixing and squeezing of parametric amplification \cite{gerryKnight2005QO} to the Hawking effect is also useful \cite{leonhardt2010essential}. The origin of the two phenomena is completely different, and even in the optical analogue, simple down conversion obey different phase matching conditions that conserve total lab frequency \cite{boyd2003nonlinear,Agrawal_NLFO} and cannot explain Hawking radiation. However, the two share intuition and the Bogoliubov transformation between incoming and outgoing modes. In both cases a pump creates 'signal' and 'idler' waves: the vacuum spontaneously generates quantum emissions, while stimulating the effect with classical light amplifies the effect for specific modes.

\section{Observing stimulated Hawking radiation}
\begin{figure}[!]
	\centering\includegraphics[width=0.9\textwidth]{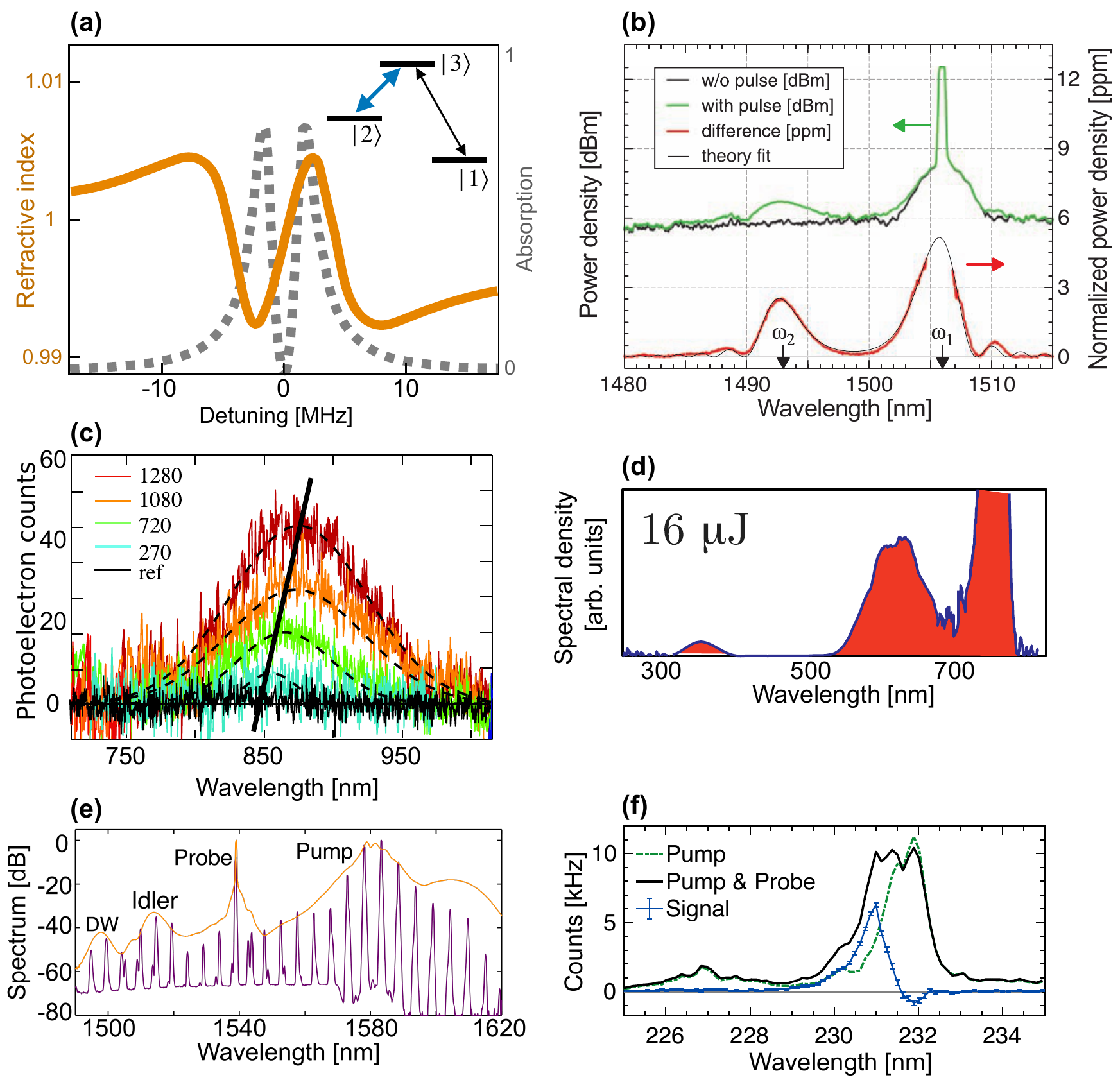}
	\caption{Central measurements. \textbf{(a)} Schematic absorption (dotted line) and refractive index (solid line) for generating 'slow light' using electromagnetically induced transparency (EIT), as the inset depicts. The refractive index has small but rapid variation at the narrow resonance. The slow group velocity is inversely related to this steep change \cite{hau1999Slow_light_1, kash1999ultraslow}. \textbf{(b)} Frequency blue-shift of probe light at fibre optical white-hole event horizon. \SI{70}{\femto\second} pump solitons were used in \SI{1.5}{\meter} long nonlinear PCF. A small fraction of CW probe at $\omega_1$ is converted to $\omega_2$, conserving its co-moving frequency, $\omega'_\text{probe}$. From \cite{philbin2008fiber}; reprinted with permission from \href{https://doi.org/10.1126/science.1153625}{\textcopyright2008 AAAS}. \textbf{(c)} Horizonless emissions from a Bessel filament creating a super-luminal refractive index perturbation in bulk silica \cite{belgiorno2010hawking, petev2013blackbody}. The radiation was seen around the phase velocity horizon (where $\omega'=0$) and increased with pump energy (in \SI{}{\micro \joule}). The wavelength shift (black line) and broadening was in agreement with the increased nonlinear refractive index. From \cite{belgiorno2010hawking}; reprinted with permission from \href{https://doi.org/10.1103/PhysRevLett.105.203901}{\textcopyright2010 APS} \textbf{(d)} Negative-frequency resonant radiation (NRR) in bulk $\text{CaF}_2$. \SI{16}{\micro\joule} Bessel pump beam at \SI{800}{\nano\meter} created dispersive-waves (around \SI{600}{\nano\meter}) and NRR (around \SI{350}{\nano\meter}) under highly non-adiabatic temporal evolution. From \cite{rubino2012negative}; reprinted with permission from \href{https://doi.org/10.1103/PhysRevLett.108.253901}{\textcopyright2012 APS}. \textbf{(e)} Comparison of frequency translation induced by a pump soliton (orange curve) and a pair of quasi-CW fields (purple curve). Probe light is being blue-shifted ('Idler') in a similar way in both cases, relating the effect to cascaded four-wave mixing. Similar dispersive-waves (DW) are also seen. From \cite{webb2014nonlinear}; reprinted with permission from \href{https://doi.org/10.1038/ncomms5969}{\textcopyright2014 Springer Nature}. \textbf{(f)} Negative frequency stimulated Hawking radiation. A compressing high-order few-cycle soliton \cite{drori2019observation} generated horizons and NRR (green dotted-dashed curve). When a pulsed probe at \SI{1450}{\nano\meter} interacted with the black-hole horizon, it red shifted (not shown) and produced negative Hawking radiation that overlapped the NRR (black solid curve). Subtracting the NRR revealed the UV Hawking signal (blue curve with 2$\sigma$ error bars). From \cite{drori2019observation}; reprinted with permission from \href{https://doi.org/10.1103/PhysRevLett.122.010404}{\textcopyright2019 APS}.
	} 
	\label{fig3}
\end{figure}

Spontaneous Hawking radiation originate from the quantum vacuum. Classical laser fields can replace the vacuum and stimulate the effect. Drori et al. \cite{drori2019observation} observed stimulated negative Hawking radiation in an optical analogue (Fig.~\ref{fig3}f). The idea of \cite{philbin2008fiber} was used, but with more suitable pump pulses and a pulsed probe. Analysis of the fibre dispersion (schematically seen in Fig.~\ref{fig2}) and simulations of the experiment allowed to optimise the experimental parameters. The pump was a high order few-cycle soliton that compressed and collapsed due to soliton fission \cite{Agrawal_NLFO}. Its peak power reached a few hundred kilo-watts, with $\SI{800}{\nano\meter}$ central wavelength. Very non-adiabatic dynamics generated NRR in the mid UV (Fig.~\ref{fig3}f), at the negative of the pump co-moving frequency, $-\omega'_\text{pump}$. This ensured the formation of steep refractive index variations, increasing the analogous surface gravity and reducing the probability of probe tunnelling. The probe was derived from the pump's laser, and was generated through cascaded Raman scattering in a meter long PCF. Negative prechirp was used to enhance the Raman induced frequency shift \cite{Rosenberg2020RIFS}, whose wavelength was continuously tuned up to $\SI{1650}{\nano\meter}$ by varying the input power. Peak powers were around $\SI{1}{\kilo\watt}$. 

Efficient and broad-band probe frequency shifts at the horizon accompanied negative Hawking radiation in the mid-UV, conserving the probe co-moving frequency $\omega'_\text{probe}$ and its conjugate $-\omega'_\text{probe}$, respectively. Since the fibre had strong dispersion, with different notions for phase and group velocity horizons, the spectrum was not Planckian \cite{drori2019observation, leonhardt2012analytical, bermudez2016hawking}. Varying $\omega'_\text{probe}$, both signals shifted according to theory, supporting the correct interpretation of the negative Hawking radiation. By this, also the interpretation of the NRR was supported, and its analytic theory \cite{conforti2013interaction} was generalised to include the Hawking process \cite{Raul2020inpress}. Linear relation was verified between the probe power and that of the negative Hawking signals up to a point where the signal saturated. This saturation was related to a back-reaction of the probe on the pulse, which is a prerequisite for Hawking radiation (since its energy is drawn from the pump, or the black-hole that curves the metric). It also slightly reduced the magnitude of the NRR (Fig.~\ref{fig3}f). 

The rate of spontaneous Hawking emission was estimated based on the magnitude of the stimulated effect \cite{drori2019observation}. It was found to be too low for measurement in the system used, calling to design an improved setup. The predicted spontaneous effect is minuscule and is overwhelmed by the noise in the system, which for the negative Hawking radiation is dominated by fluctuations of the overlapping NRR. The multi-mode nature of the PCF in the UV \cite{Agrawal_NLFO} was found to reduce the observed signal power by up to four orders of magnitude.

The experiment \cite{drori2019observation} showed how robust the Hawking effect is: it appeared despite the extreme nonlinear dynamics of the collapsing soliton. Since time in the co-moving frame is related to the propagation distance along the fibre, rapid variations in the pump profile do not violate the required slow evolution of the metric as long as they appear over a length scale much larger than the Hawking radiation wavelength \cite{visser2003essential}. As such, pump deceleration due to the Raman effect could be accounted by simply correcting its central frequency.

\section{Beyond the horizon?}
Analogue gravity in optics has come a long way: from erroneous visionary ideas, to careful analysis of experimental demonstrations. It drove new research directions and understandings in optics. The reality of optical horizons and negative frequencies became clear, and multiple optical interpretations made their physics more tangible. The crucial role of dispersion was stressed, determining the Hawking spectrum (Fig.~\ref{fig2}). A group velocity horizon is blocking the radiation modes and allow their efficient conversion. A phase velocity horizon marks the support of non-trivial negative frequency modes, that allow for the Hawking amplification. The optical and other analogues, especially in water waves and BECs, provide extensive insights for gravity through concrete real-world examples. The universality of the Hawking effect separates it from gravity and high-energy physics, and emphasises the central role of classical fields to the process. This hints to more possibilities in gravity \cite{barcelo2019analogue} and in optics \cite{leonhardt2015cosmology}, even before going fully quantum. 

The main challenge for observing the spontaneous (quantum) effect in optics is its low power. Noise from spurious radiation makes matters worse. Using knowledge from the stimulated effect \cite{drori2019observation} could help overcome these challenges.

The demonstrated robustness of the Hawking effect is calling for new experiments to lead the way. Our increased confidence in the interpretation of the results allows to further separate from the traditional scheme of Hawking radiation, developing bolder questions and ideas. Optical technologies allow to realise wild ideas and to study them with high precision. We call for more cross-fertilisation between research of the different black-hole analogues, which might open new horizons in these fields.

\enlargethispage{20pt}

\section{Funding}
Fellowship of the Sustainability and Energy Research Initiative (SAERI) program, the Weizmann Institute of Science; European Research Council; and the Israel Science Foundation.

\section{Acknowledgements}
I am grateful for discussions and comments from David Bermudez, Jonathan Drori and Ulf Leonhardt. I acknowledge valuable discussions with the participants of the scientific meeting 'The next generation of analogue gravity experiments' at the Royal Society (London, December 2019), and thank its organizers: Maxime Jacquet, Silke Weinfurtner and Friedrich K{\"o}nig.

%%%%%%%%%% Insert bibliography here %%%%%%%%%%%%%%

\bibliography{RSReview} 

\begin{thebibliography}{100}
\bibitem{HawkingTemp}
The temperature of Hawking radiation from a black-hole of solar mass is about
  \SI{60}{\nano\kelvin}, where CMB temperature fluctuations are about
  \SI{100}{\micro\kelvin}.
\end{thebibliography}
\bibliographystyle{ieeetr}

\end{document}